\pdfoutput=1
\documentclass[aps,pre, twocolumn, groupedaddress]{revtex4-1} 
\usepackage{lipsum}
\usepackage{mathtools}
\usepackage{graphicx}
\usepackage{dcolumn}
\usepackage{amsmath}   % need for subequations
\usepackage{amssymb}
\usepackage{bm}
\usepackage{hyperref}
\usepackage{latexsym}
\usepackage{verbatim}
\usepackage{color}

\usepackage[caption=false]{subfig}

\def\beq{\begin{equation}}
\def\eeq{\end{equation}}

\def\beq{\begin{equation}}                          
\def\eeq{\end{equation}}                          
\def\bea{\begin{eqnarray}}                          
\def\eea{\end{eqnarray}}

\DeclareRobustCommand{\uvec}[1]{{%
  \ifcsname uvec#1\endcsname
     \csname uvec#1\endcsname
   \else
    \bm{\hat{\mathbf{#1}}}%
   \fi
}}

\preprint{}

\begin{document}

%\preprint{}

%%%%%%%%%%%%%%%%%%%%%%%%%%%%%%%%%%%%%%%%%%%%%%%%%%%
%                               TITLE & ABSTRACT
\title{Macro to micro phase separation in a collection of chiral active swimmers}
\author{Vivek Semwal}
\email{viveksemwal.rs.phy17@itbhu.ac.in}
\affiliation{Indian Institute of Technology (BHU) Varanasi, India 221005}
\author{Jayam Joshi}
\email{jayam.joshi.phy20@iitbhu.ac.in}
\affiliation{Indian Institute of Technology (BHU) Varanasi, India 221005}
\author{Shradha Mishra}
\email[]{smishra.phy@itbhu.ac.in}
\affiliation{Indian Institute of Technology (BHU) Varanasi, India 221005}
\date{\today}
\begin{abstract}
We studied a collection of chiral active particles (CAP) on a two dimensional substrate using extensive numerical study. Particles interact through soft repulsive interaction. The activity and chirality of particles is tuned by varying their self-propulsion speed and angular velocity respectively. Kinetics and steady state properties of particles are studied for different chirality and activity. The phase diagram of system on the plane of activity and chirality shows  three distinct phases. For small chirality when activity is dominant, particles show enhanced dynamics and  macroscopic phase separation of ordered clusters is observed. For moderate chirality, micro clustered phase 
is observed in which small clusters with moderate ordering are formed. For large chirality, when chirality dominates, no clustering is found because particle motion is mainly confined to its location. Our study gives a detail insight into the  effect of chirality on the properties of collection of CAP, which can be useful to understand the dynamics and steady state of many natural micro swimmers.
 
\end{abstract}
\maketitle
\section{Introduction}
%%%%%%%%%%%%%%%%%%%%%%%%%%%%%%%%%%%%%%%%%%%%%%%%%%%%%%%%%%%%%%%%%%%%%%%%%%%%%%%%%%%%%%%%%%
Active Brownian particles (ABPs) are  prominent example of active matter  \cite{sriram2005,sriramflock,ghosh2009controlled,liao2018clustering,levis2018micro}. ABPs combine Brownian motion with self-propulsion. Motile microorganisms are frequently characterised as ABPs, in addition to artificial self-propelled microparticles \cite{bechinger,liao2018clustering,levis2018micro}. Even bacteria that conduct a run-and-tumble action \cite{Peruani_2013,bechinger,enhance1,enhance3}, such as Escherichia coli, have been effectively classified as ABPs. One of the remarkable property of ABP's are motility induced phase separation (MIPS) without any cohesive interaction among the particles \cite{mipsc,mips2013prl,bechinger,enhanced,enhancediff,enhancediffusivity}. Most of theoretical and simulation study of ABP is focused on  systems without chirality \cite{mipsc}. But chirality is an inherent property in many natural active particles  \cite{bechinger,elgeti2010hydrodynamics,menzel2015tuned,kraft2013brownian,ghosh2009controlled,liao2018clustering}. Hence effect of chrality on the properties of ABPs is an important question to be asked. 

In \cite{keaveny2009hydrodynamic,keaveny2013optimization}, motion of microswimmers  is studied in the presence of chirality. Chirality leads to the deviation of particle trajectory from the straight line motion. In a recent review \cite{lowen2016chirality}, it is shown that an individual or collection of chiral or  circle swimmer can show interesting properties. When present in bulk they can also 
show active turbulence \cite{keaveny2013optimization,winkler2016low,ginelli2016physics,levis2018micro}. Our study is motivated with recent study of 
\cite{ma2022dynamical} dynamic clustering of chiral active particles. Chirality of particles suppress the motility induced phase separation present for nonchiral active Brownian particles \cite{cates2015motility}. 
Hence effect of chirality on the properties of individual and collective behaviour of active particles can give good understanding of another class of nonequilibrium system called chiral active particles (CAPs). Most of the recent study of CAPs have focused on the effect of chirality on the kinetics or steady state properties of active particles on the variation of activity or packing densities of the particles \cite{ghosh2009controlled,liao2018clustering,levis2018micro}.  
The study of properties of active particles, on the variation of chirality is very scarce. But such study can provide a good understanding of
effect of chirality on the properties of active particles. The two extremes: small or large chirality is trivial: for the first case we expect the results of ABPs in the collection and for the later case we expect  mainly the confined circular motion. But what happen when we slowly tune the chirality from small to large values still unexplored. In the present work we focus on this effect of variation of chirality on the properties of active particles.

Here we show the kinetics and steady state properties of CAPs on the variation of chirality and activity. The system is found in three distinct phases: (i) for small chirality, when activity dominates, system shows the enhanced diffusion\cite{ma2022dynamical,PhysRevLett.126.188002,klamser2018thermodynamic} and macroscopic clustering as found in MIPS. In the second phase where both activity and chirality are in competition, we  find some clustering, but no macro phase separation. For  larger chirality, the chirality dominates over activity and dynamics of particles is mostly confined to its location and no clustering is observed.

Our article divided in the following sections: In section \ref{secI}, we give the detailed description of our model. In section \ref{secII} discuss about the results of numerical simulation of the system. In section \ref{secdis} we conclude our result and discussion about the future directions of our study.
%%%%%%%%%%%%%%%%%%%%%%%%%%%%%%%%%%%%%%%%%%%%%%%%%%%%%%%%%%%%%%%%%%%%%%%%%%%%%%%%%%%%%%%%%%
\section{Model}\label{secI}
Our system consists of $N$ chiral  active particles (CAP) of radius $a_0$ on a two-dimensional substrate. 
On the substrate each $i^{th}$ particle is represented by its position  vector
${\bf r}_{i}(t)$ and orientation $\theta_i(t)$,  at time $t$.  
The dynamics of the particle is governed by the overdamped Langevin equation \cite{vivek,langevin,langevin2,noise,semwal2021dynamics}
\begin{equation}
	\partial_t{\bf{r}}_i=v{\bf \hat{{n}}_i}+\mu_1\sum_{j\neq i}{\bf {F}}_{ij}
\label{eq(1)}
\end{equation}
 \begin{equation}
	 \partial_t\theta_i=\omega +\sqrt{2D_{r}}\bf{\eta _{i}}
\label{eq(2)}
\end{equation}

\begin{figure} [hbt]
{\includegraphics[width=1.0 \linewidth]{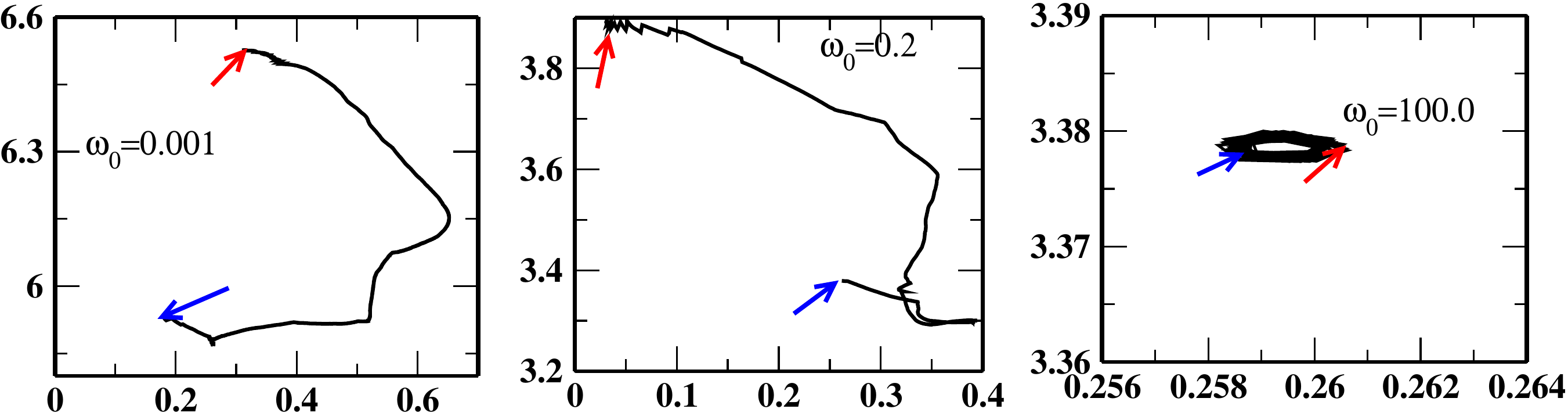}}
\caption{(color online) Trajectory of the particle for (a) small $\omega_0=0.001$, (b) intermediate $\omega_0=0.2$ and (c) large $\omega_0=100$. The blue and red arrows show the starting and end of the trajectory for a fixed time interval. The numbers on the axes show the coordinates of the particle at different times. Notice the change in the extend of the trajectory for three different $\omega_0$'s. }

\label{fig:1}
\end{figure}
The first term on the right hand side (RHS) of Eq. \ref{eq(1)} is due to the 
activity of the particle, and $v$ is its self-propulsion speed. The particle moves along its unit orientation
direction vector $ {\bf \hat{n}}_i = (\cos{\theta_i},\sin{\theta_i})$ with speed $v$. The second term represents
the steric force, ${\bf F}_{ij}=F_{ij}{\bf \hat{ r}}_{ij}$, which takes care of the repulsive 
 interaction, acting on the $i^{th}$ particle  due to its neighbouring particles in contact with it. Hence we consider $F_{ij}=k(2 a_0-r_{ij})$ 
if $r_{ij} \leq {2a_0}$ and $F_{ij}=0$ if $r_{ij} \geq {2a_0}$,
where $r_{ij}$ is the centre to centre distance between $i^{th}$ 
and $j^{th}$ particles, $r_{ij}=\vert{{\bf r}_i-{\bf r}_j}\vert$. $k$ is the strength of the force,  
$F_{ij}$ is tuned by the mobility $\mu_1$ of the particles. Further, the orientation of the particle is updated by Eq. \eqref{eq(2)} 
where $\omega$ is the chirality (angular velocity) of the particle and $\eta_{i}$ is Gaussian white noise term. 
 $D_{r}$  is the rotational diffusivity.
The smallest time step 
considered is $\delta t=0.005$. We define the dimensionless chirality  as $\omega_{0}=\frac{\omega}{D_r}$,   and  the dimensionless activity of the particles  as $v_{0}=\frac{v}{a_0\times D_{r}}$. 
We start our simulation with random position and orientation of particles on the substrate and evolve 
the system by integrating the Eqs. \ref{eq(2)} and \ref{eq(2)} using Euler's 
integration scheme. The system is simulated for total simulation time steps of $10^6$. One simulation step
is counted after update of all the particles once. All the physical quantities calculated here are averaged $50$ realizations. 
The tuning parameters are chirality $\omega_0$ and activity $v_0$. We tuned $\omega_0$  from $0$ to $100$ and $v_0$ from $1$ to $10$.
Simulation is performed in box of size $L=150\times a_{0}$, with packing fraction  $\frac{N \pi \sigma^{2}}{L^{2}}=0.6$.

{\section{Results}\label{secII}}
We first observe the  trajectory of a single  particle in the collection for different chirality. In Fig. \ref{fig:1}(a-c) we show the 
trajectory for three different values of chirality  $\omega_{0}=0.001$, $0.2$ and $100$ respectively and for fixed self-propulsion speed $v_0=10$. 
For small $\omega_{0}=0.001$, Fig. \ref{fig:1}(a) trajectory looks very
extended. The blue and black arrows show the starting and end point of trajectory respectively. For 
intermediate $\omega_{0}=0.2$, Fig. \ref{fig:1}(b), trajectory looks localised for early time and then extended towards the end. For large $\omega_{0}=100$, Fig. \ref{fig:1}(c)
trajectory is always confined within a small region. For comparison the time difference from the start and end of the trajectory is kept the same for 
all three cases.\\ 

\begin{figure*}[hbt]
        {\includegraphics[width=1.0 \linewidth] {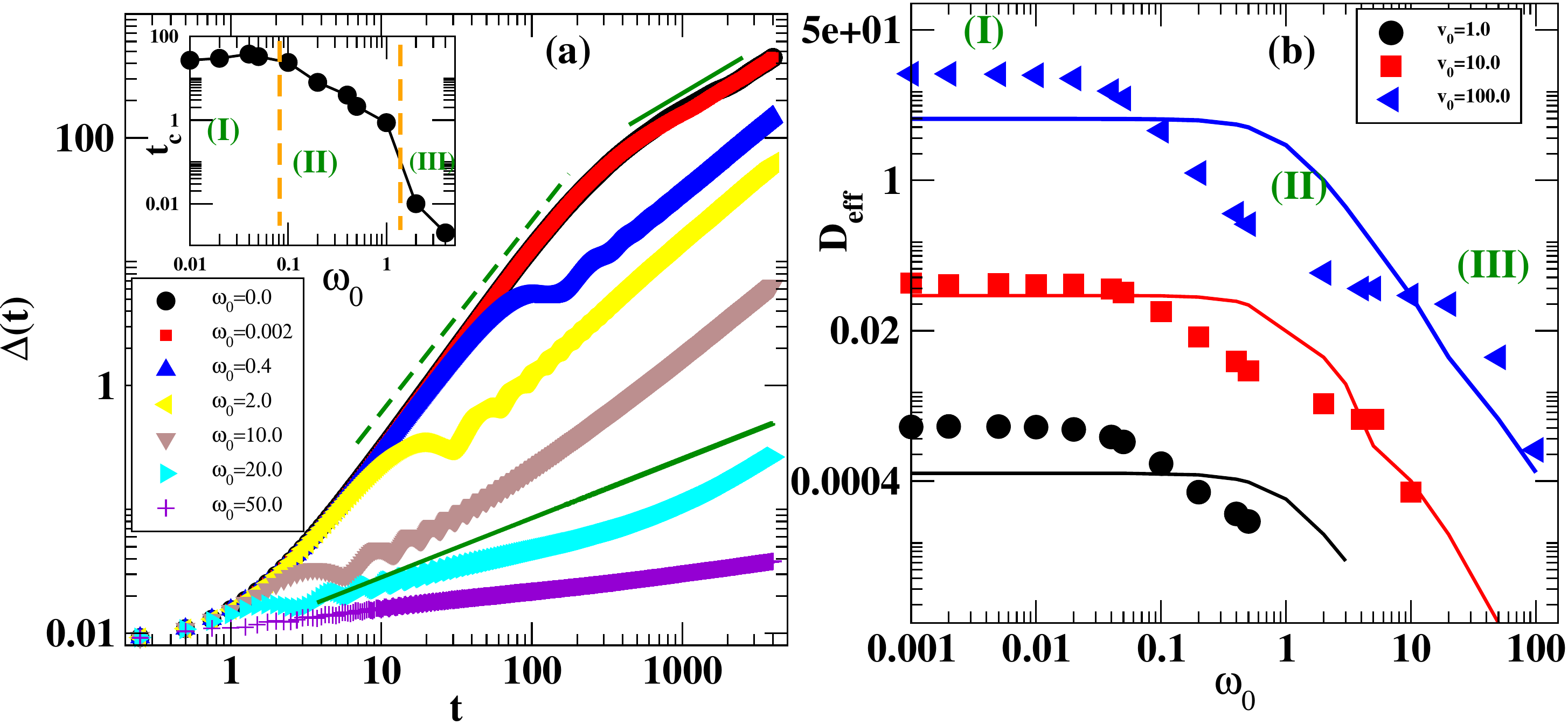}}
        \caption{(color online) (a) Plot of MSD, $\Delta$(t) vs. time $t$  for different value of chirality. Dashed and solid lines are of slope $2$ and $1$ respectively. (inset)  crossover time $t_c$ vs. $\omega_{0}$. Dotted orange lines show three different regime I, II and III. (b) $D_{eff}$ vs. $\omega_{0}$ for three different values of $v_0$. Different symbols  show numerical data, straight line shows the plot of Eq. \ref{analytical1}}
	\label{fig:2} 
\end{figure*}

To quantify the above observation, we measure the mean square displacement of the particles for  different  chirality.
We define the particles mean square  displacement as $\Delta (t) =<(r_i(t+t_0)-r_i(t_0))^2>$ where $<..>$ denotes average over all the particles, many reference times $t_0$ and over different realisations. In general the dynamics of active particles shows an early time ballistic dynamics and then crossover to late time diffusion. In Fig. \ref{fig:2} we show the behaviour of MSD for fixed activity $v_0=10.0$ and varying the chirality. For zero chirality 
system shows a very clear crossover from early time ballistic to late time diffsuion. As we introduce chirality,  $\Delta(t)$ shows oscillations, which is due to the oscillations of particles trajectory for finite chirality. The periodic oscillations increases and crossover time decreases on 
increasing $\omega_0$. We extract the typical crossover time $t_c(\omega_0)$ by fitting the MSD with the persistent random walk= 
{\bf $\Delta(t)=4D_{eff}t [1-exp(-t/t_{c})]$}, for different $\omega_{0}$ and plot is shown in the inset of Fig. \ref{fig:2}(a). The crossover time remains almost constant for smaller chirality $\omega_0\le0.1$ and then show a smooth decay for intermediate $0.1<\omega_0<10.0$ and then decays sharply for larger $\omega_0>10.0$. For large chirality the dynamics of particle is no longer diffusive for late time and hence $t_c$ cannot be calculated. \\
To further characterise the dynamics of particle we also calculated the late time effective diffusivity $D_{eff}(\omega_0)$ for different $\omega_0$ 
for three different $v_0 = 1, 10$ and  $100$. The late time effective diffusivity $D_{eff}$ is obtained by
$D_{eff} = \lim_{t \to \infty} \frac{\Delta(t)}{4t}$ 
%\underset{t \to \infty  }{=}\frac{\Delta (t)}{4t}$. 
The plot of $D_{eff}(\omega_0)$ vs. $\omega_0$ is shown in Fig. \ref{fig:2}(b). For large activities $v_0=10$ and $100$, we find that
for small chirality $\omega_0\le0.1$, the $D_{eff}(\omega_0)$ remains flat and then for the intermediate $\omega_0 \in (0.1, 10)$,  shows a 
shallow region with slow decay and finally for larger chirality decay sharply to very small values. For smaller activities $v_0=1.0$, the dynamics for larger chirality is mostly confined hence the $D_{eff}$ cannot be defined.\\
For comparison we also calculated the MSD for noninteracting single chiral particle analytically. The calculation is performed by making the interaction force term zero. 
In the absence of interaction, the overdamped Langevin equations \ref{eq3} and \ref{eq4} reduces to:
\begin{equation}
	\partial_t{\bf{r}}_i=v{\bf \hat{{n}}_i}
	\label{eq3}
\end{equation}
 \begin{equation}
	 \partial_t\theta_i=\omega +\sqrt{2D_{r}}\bf{\eta _{i}}
	 \label{eq4}
\end{equation}
these equations can be solved to obtain the mean square displacement of chiral active particle without interaction. The first and second moments of $ \theta(t) $ are simply $<\theta(t)> = \theta_0 + \omega t $ and $ <\theta^2(t)> = \omega^2 t^2 + 2 D_r t $, where, $\theta_0 = \theta(t=0) = 0 $. Using this, the second moment of $r(t) $ or the mean square displacement is obtained: 
\begin{multline*}
    <r^2(t)> = \Delta(t) = \frac{2v^2}{(\omega^2 + D_r^2)^2}[\omega^2 - D_r^2 + D_r(D_r^2 + \omega^2)t \\
    + e^{-D_rt}((D_r^2 - \omega^2)cos(\omega t) - 2D_r\omega sin(\omega t))]
\end{multline*}
The late time diffusivity or $ D_{eff}$ is defined to be $\lim_{t \to \infty} \frac{\Delta(t)}{4t}$.  Hence in the absence of interaction the diffusivity is given by the expression:
\begin{equation}
    D_{eff} = \frac{v^2 D_r}{2(\omega^2 + D_r^2)}
    \label{analytical}
\end{equation}
Writing $D_{eff}$ in terms of dimensionless chirality $\omega_0$,
\begin{equation}
    D_{eff} = \frac{v_{0}^2}{2D_{r}(\omega_{0}^2 + 1)}
    \label{analytical1}
\end{equation}
In Fig. \ref{fig:2}(b),  lines are from the analytical expression of $D_{eff}$ as given in Eq. \ref{analytical1}. For small chirality for all
activities the $D_{eff}$ for noninteracting case is smaller than the interacting full numerical simulation. Later we are going to show for the same range of chirality particles
show the clustering. Hence in this regime collective dynamic of particles is responsible for the enhanced dynamics in comparison to the noninteracting single particle dynamics. 
As we increase chirality, the noninteracting $D_{eff}$ decay smoothly to zero values for all activities, whereas for interacting case, it shows a small plateau for intermediate 
chirality. Hence we  can say that the second region with intermediate chirality is all due to the interaction between the particles. Later we are going to explore the system more for 
the three different regions using numerical study of interacting system. Further we classify the three regions for the  interacting system as region $I$, $II$ and $III$ as marked 
in Fig. \ref{fig:2}(b). In next sections we discuss in detail the how does the clustering of particles is changed due to the effectively three different dynamics in the three regions.
\\
\subsection{Cluster size distribution}
\begin{figure*}[hbt]
\centering
\begin{tabular}{cc}
\includegraphics[width=0.5\textwidth]{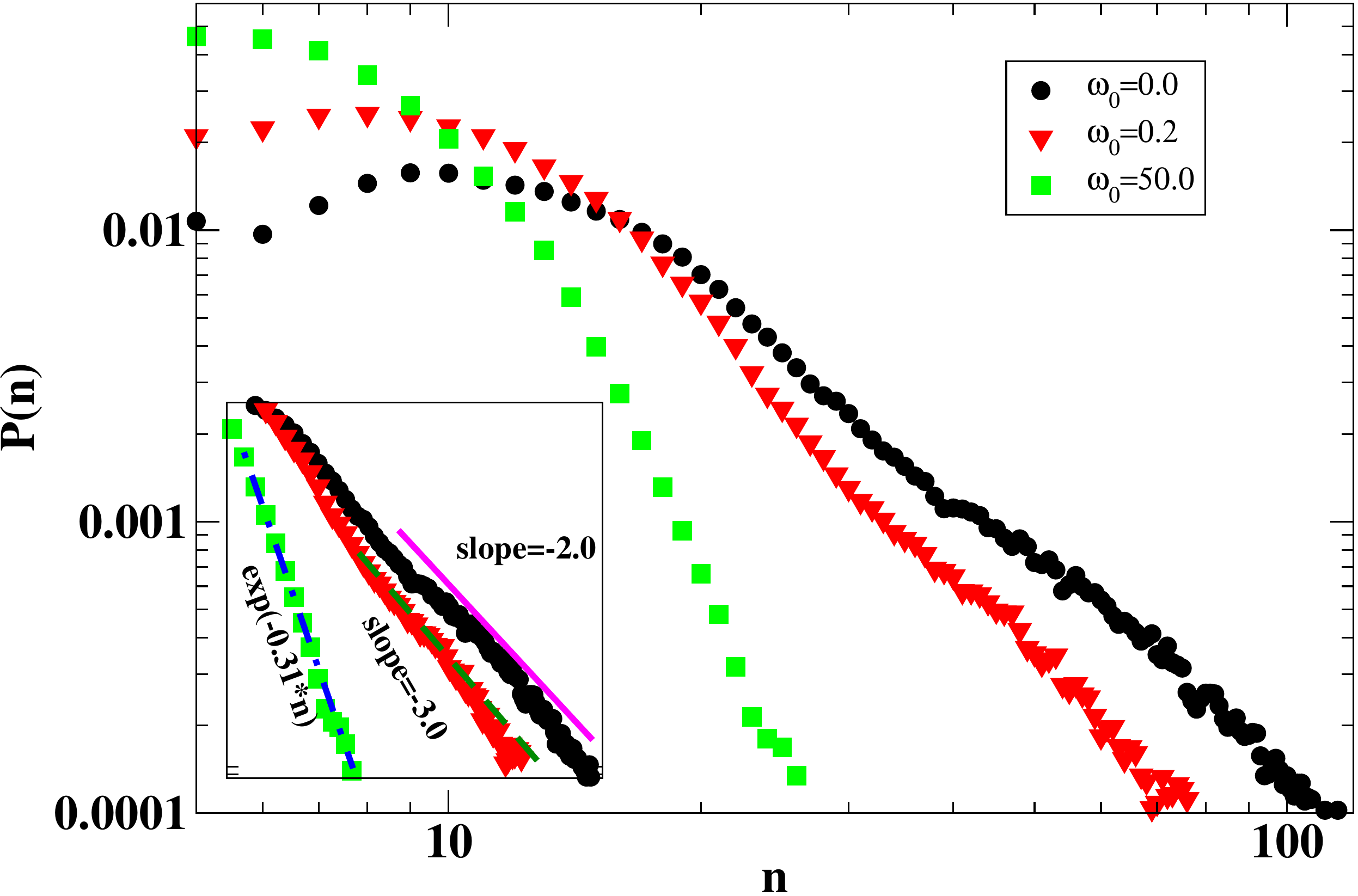} & \includegraphics[width=0.43\textwidth]{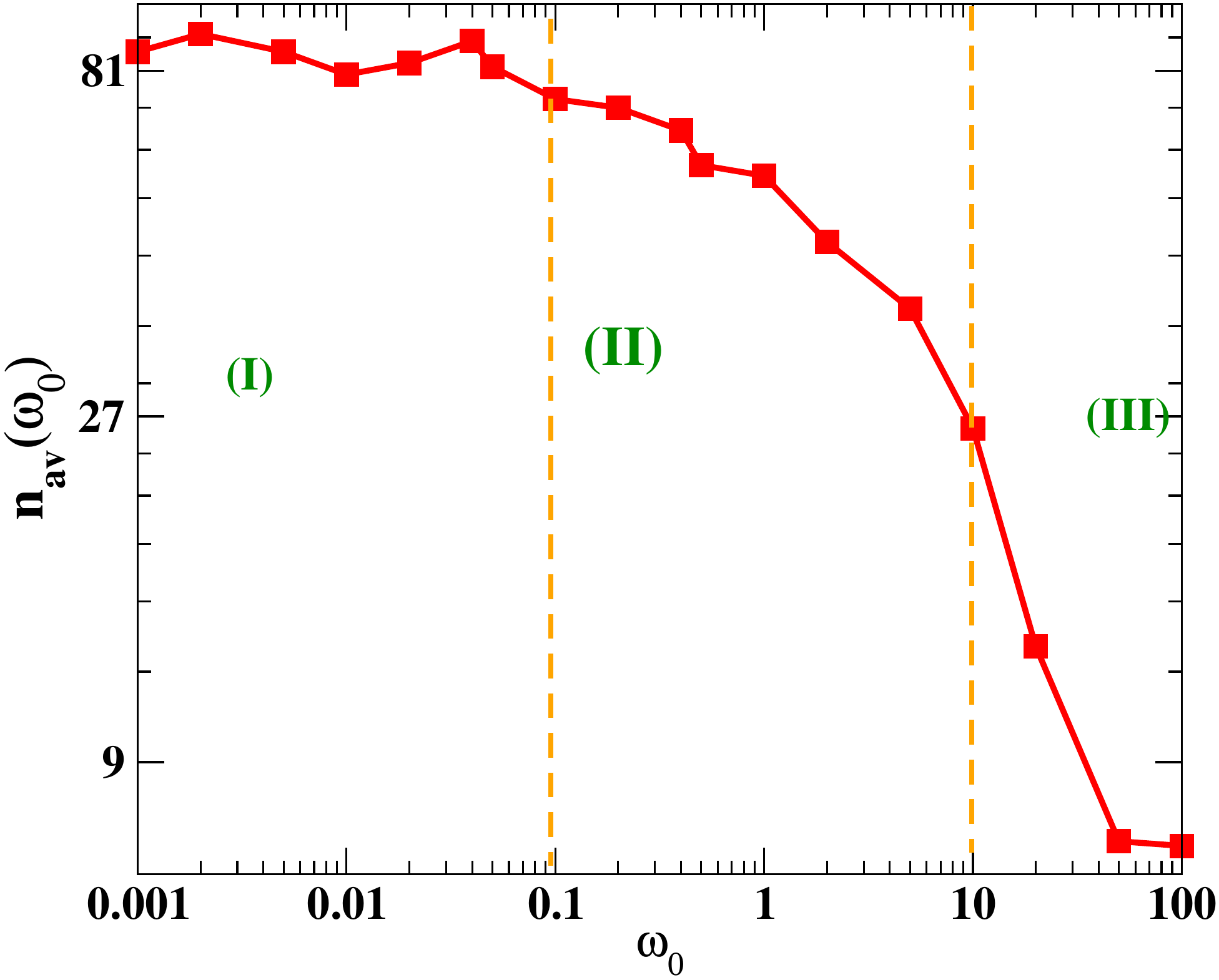} \\
(a)&(b)
\end{tabular}
\caption{(a) $P(n)$ vs.  $n$ for $v_{0}=50$ for three different values of $\omega_0 = 0, 0.2$ and $50$ chosen in three different regions. Inset: shows the zoomed plot of $P(n)$ vs. $n$ for large $n$. The solid  and dashed lines are lines of slope $-2$ and $-3$ respectively. The dotted dashed curve is exponential fitted curve.
(b) $n_{av}( \omega_{0} )$ vs. $\omega_{0}$. Dotted orange lines show the different regions I, II and III.}
\label{fig:3}
\end{figure*}

\begin{figure*}[htb]
\centering
\begin{tabular}{ccc}
\includegraphics[width=0.32\textwidth]{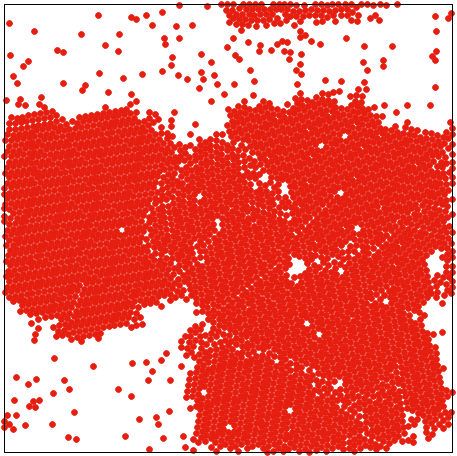} &\includegraphics[width=0.32\textwidth]{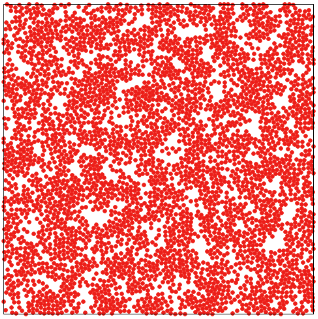}&\includegraphics[width=0.32\textwidth]{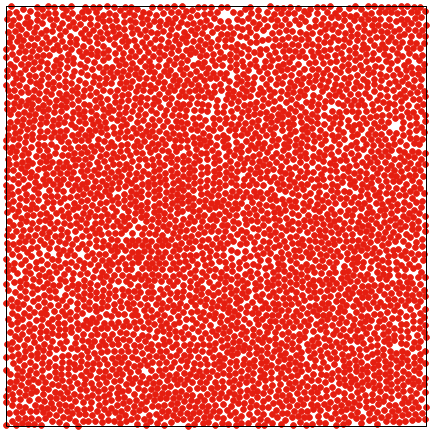}\\
(a) & (b) & (c)\\
\end{tabular}
\caption{Snapshots in three different regions (a) for $\omega_{0}=0.001$,  (b) for $\omega_{0}=1.0$ (b) and (c) for $\omega_{0}=50$ and all are for $v_{0}=10$ .}
\label{fig:4}
\end{figure*}

{\label{clustersize}}
In the previous section we studied the effect of chirality on the effective dynamics of the system. Now we study how 
the change in effective dynamics changes the clustering  and phase behaviour of particles in the system.
In Fig. \ref{fig:4}(a-c) we show the snapshots of particles for three different $\omega_{0}$ $=0.001, 1.0$ and $50$ respectively in regions
I, II and III for activity $v_0=10.0$. For the I region, Fig. \ref{fig:4}(a) we clearly see the 
macroscopic clustering in the system,   
As we to to the region II, Fig. \ref{fig:4}(b) we see the microscopic clustering,
Finally in region III, Fig. \ref{fig:4}(c) there is no clustering and system is homogeneously distributed.
Further we 
calculate the cluster size distribution (CSD) for different chirality for activity $v_0=10$ as shown  
in Fig. \ref{fig:3}(a). The CSD is defined using the particles connected by a most probable distance $r_{0}$. In Fig. 
\ref{fig:3}(a) we plot the normalised CSD $P(n)$ vs. $n$ for different chirality $\omega_{0}$, specifically chosen in 
the three regions of the $D_{eff}$ plot shown in Fig. \ref{fig:2}(b), where
$n$ is the size of the cluster. We find that for small $\omega_{0} = 0.2$, 
$P(n)$ decays as power law $\simeq \frac{1}{n^{\alpha}}$ for large $n$, with exponent $\alpha=2$ as reported in 
previous studies \cite{PhysRevLett.96.204502}. As we increase $\omega_{0}$ and system transits into the second phase 
$P(n)$  still  decay as  power law, but the exponent $\alpha \simeq 3$, On further increasing the chirality of the particle $P(n)$ decays exponentially with $n$. Hence the three regions which are defined based on the effective
dynamics of particles in the steady state: also lead to  different types of clustering of particles. We also calculated the average cluster size $n_{av}(\omega_{0})$ for different $\omega_{0}$ and found that it also shows  three different regions as shown in Fig. \ref{fig:3}(b) for $v_0=10$.
$n_{av}(\omega_{0})$  is defined as $n_{av}(\omega_{0})=\int n P(n) dn$. 
In the first region as shown in Fig.{\ref{fig:3}(b)}, system shows the formation of macroscopic clusters with an average cluster 
size $n_{av}(\omega_{0})$ around 75 in the (I) phase,  $n_{av}(\omega_{0})$  varies from $70$ to $25$ for the $\omega_{0}$ in the (II) phase. As we 
further increase $\omega_{0}$, system enters in to the third region and homogeneous state of the particles and $n_{av}$ decay sharply  to very small values.
Now we try to understand how the chirality affects the structural ordering of particles in the system.
\subsection{Structural ordering} 
\begin{figure*}[hbt]
\centering
\begin{tabular}{cc}
\includegraphics[width=0.5\textwidth]{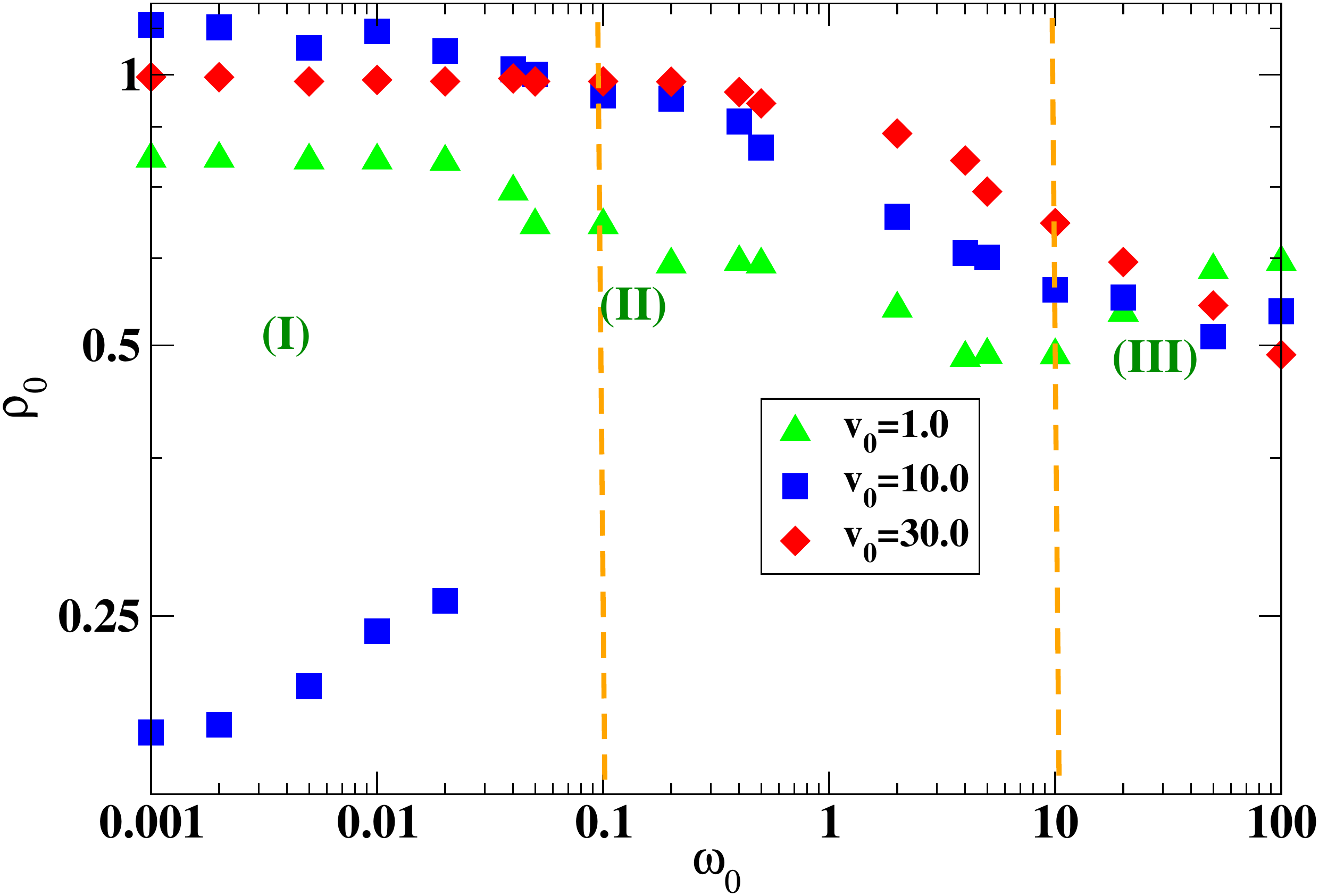}&
  \includegraphics[width=0.5\textwidth]{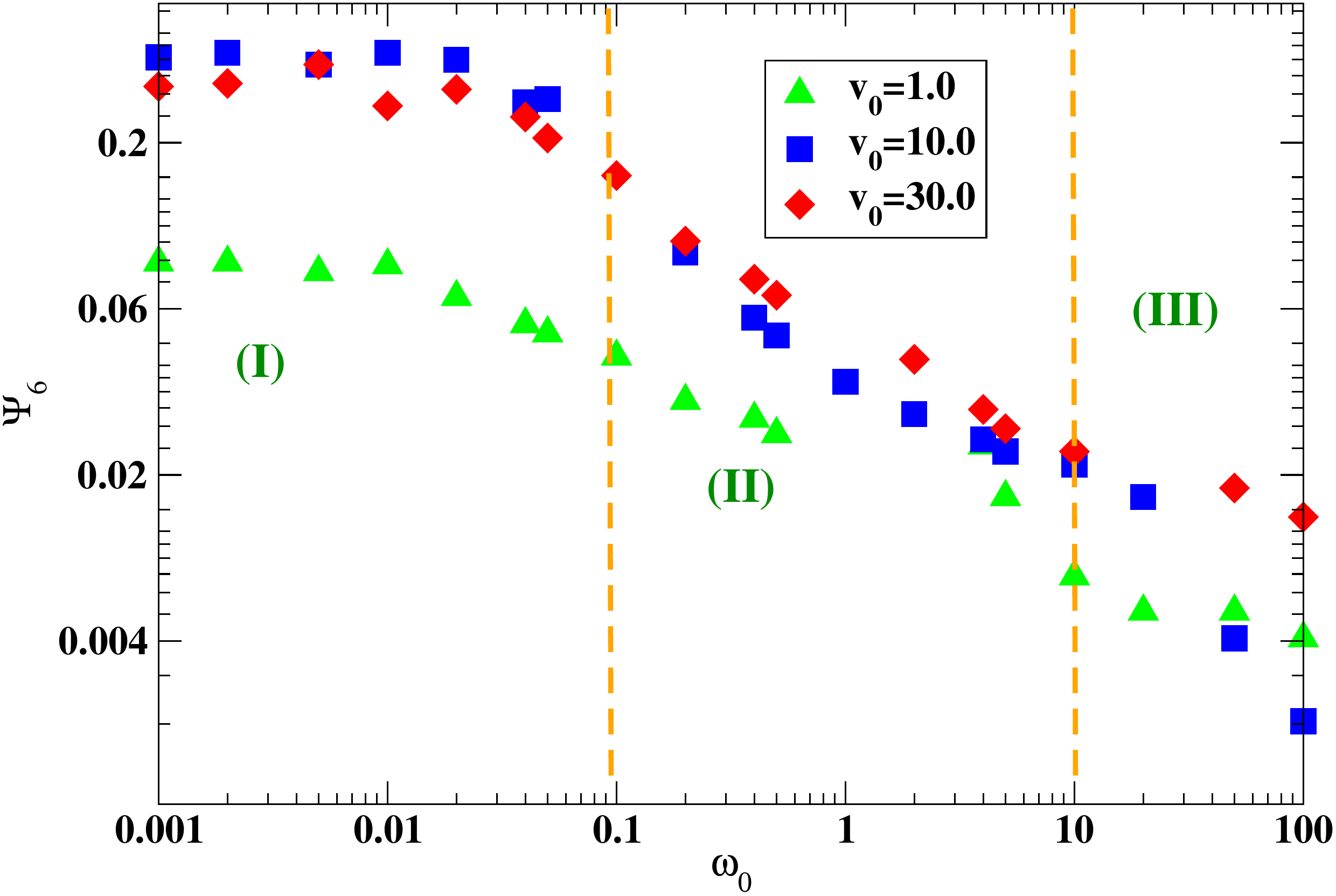}\\
(a) & (b)
\end{tabular}
\caption{(a) $\rho_0$ vs. $\omega_{0}$  for different values of $v_{0}$. . (b) Plot of $ \Psi_{6}  $ vs. $\omega_{0}$ for different values of $v_{0}$.
Dotted orange lines have the same meaning as in Fig. \ref{fig:3}}
\label{fig:6}
\end{figure*}

{\label{si6}}
We further try to understand the effect of chirality on phase separation and structural ordering in the system.
 We define local density 
around a particle $\rho_{loc}$ as function of  chirality for different activities. In general a nonchiral system 
shows a motility induced phase separation (MIPS) on tuning the activity or packing fraction \cite{klamser2018thermodynamic}. 
Here we are interested 
on the effect of chirality keeping packing fraction fixed and for different activity. We define the 
local density $\rho_{loc}$ with the help of the number of particles surrounding a given particle.  $\rho_{loc} = \frac{n_{p}}{6}$, where $n_p$ is the number of particles surrounding 
the given particle. For a perfectly packed 
surrounding we expect number is $6$ for hexagonal close packed (HCP) structure and then we define that 
the $\rho_{loc}=1$.  Hence in this way for a given snapshot of the system we have a distribution of $\rho_{loc}$, $P(\rho_{loc})$
where the probability distribution function $P(\rho_{loc})$ is obtained by looking $\rho_{loc}$ of each particle.
For a perfect clustered phase the $\rho_{loc}$ will be peaked  around $1$ for completely
homogeneous phase $\rho_{loc}$ will approach the mean packing density $\rho_{loc}=0.6$ of the system.
In Fig. \ref{fig:6}(a) we show the plot of 
location of peaks $\rho_0$ of $P(\rho_{loc})$ vs. $\omega_{0}$ for three different values of activities 
$v_0=1,10,30$. We find that for small $v_0=1$, $P(\rho_{loc})$ has only one peak. For small $\omega_{0}$, $\rho_0$ remains flat close to $0.7$ and then decay to mean packing density $0.6$ for larger chirality. For  activity $v_0=10$, the $P(\rho_{loc})$ is bimodal (data not shown) and location of two peaks at 
smaller and larger $\rho_0$'s is shown in Fig. \ref{fig:6}(a). For small chirality or in the I region
the two peaks are widely separated and as we enter the II region the difference between two peaks 
diminishes and for IIIrd region or for high chirality $\rho_0$ approaches value close to mean packing density. For larger activity $v_0=30$, in the I region $\rho_0$ remains close to 1 and the smoothly decay to moderate values in the II region and finally approaches to $0.6$ in the III region. For large activities, we do not find bimodal distribution of $P(\rho_{loc})$ due to the very strong clustering in the system. \\ 
   
   Now to further understand the effect of chirality on the structural ordering in the system.
   
We define the local bond order parameter $\psi_6(t)$ \cite{mermin1968crystalline,lechner2008accurate}
\begin{equation}
\psi_6(t) = \frac{1}{N_p} \sum_{k=1}^{N_p} \sqrt{\frac{1}{N_k}\sum_{j=1}^{N_k}e^{i6\theta_{kj}}}
\end{equation}

 $\psi_6(t)$ measures the amount of  hexagonal ordering in the system. For the perfect  hexagonal close packed (HCP) structure, $\psi_6(t)$ will be close to  $1$ and for perfect 
random arrangement it is close to $0$. In Fig. \ref{fig:6}(b) we show the plot of mean 
value of $\psi_6(t)$, $\Psi_6(t) = <\psi_6(t)>$, where the mean $<..>$ is average over time in 
the steady state and over realisations. For  $v_0=1$, $\Psi_6$ remains close to small values $0.08$ for small chirality and decay on increasing $\omega_{0}$. For large activity and small chirality (region I) the $\Psi_6 \simeq 0.65$ and remains flat in the first region and then smoothly decay in region II and finally approaches very small values of the $\mathcal{O}(10^-2)$ as system approaches III region. Hence very clearly not only density shows the three types of clustering in three regions, but structural ordering also distinctly shows  three regions with variation of chirality for high activities.\\
\section{Phase Diagram} 
\begin{figure} [hbt]
{\includegraphics[width=1.0 \linewidth]{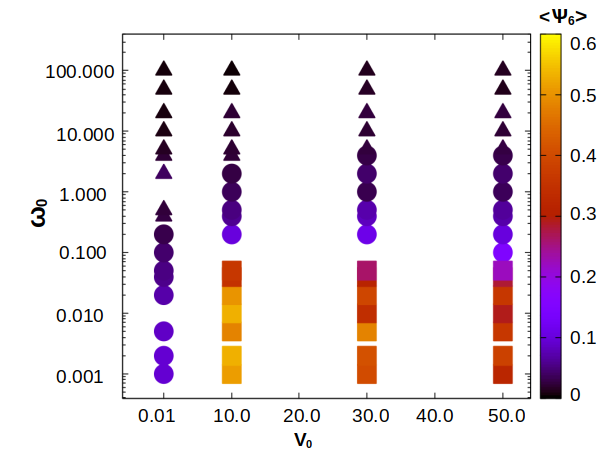}}
\caption{(color online) Phase diagram of the system in  ($\omega_{0}$, $v_{0}$) plane. Squares show the MAC phase, circles show the MIC phase and rectangle shows the HS phase.
Color bar shows the magnitude of $\Psi_6$ order parameter.}
\label{fig:5}
\end{figure}

{\label{phasedg}}
Based on the above results of $\Psi_6$ we have drawn the phase diagram in the plane of ($v_0$, $\omega_{0}$). The system is found in three distinct phases: (i) Homogeneous State (HS) defined as small structural ordering 
$\Psi_6 <0.025$, $\rho_0$ close to mean packing density $0.6$, exponential CSD shown by triangles in the figure \ref{fig:5}. This phase is found for large chirality and all activities. (ii) Microscopic cluster (MIC), defined as moderate structural 
ordering $0.025 < \Psi_6 <0.25$, $\rho_0$ decays to moderate values and CSD decay algebrically with larger exponent $3$. This phase is shown as circles in the phase diagram. (iii) Macroscopic cluster (MAC), with large $\Psi_6 >0.25$, large $\rho_0$ close to $1$ and CSD decays algebrically  with power close to $2$. This phase is shown using squares in the phase 
diagram. The color shows the value of $\Psi_6
$ for different parameters ($v_0$, $\omega_{0}$).\\
{\section{Discussion}\label{secdis}}
In conclusion we show the effect of varying the chirality on the collection of circle micro swimmers. The competition between activity and chirality
leads to three distinct phases as we slowly tune the chirality. For small chirality when linear motion dominates, effective dynamics of 
particles is enhanced, in comparison to single chiral particle with the same chirality. It leads to macroscopic clustering of particles. For  intermediate chirality when linear and circle motion are comparable, the particles show weaker clustering with small cluster formation. The effective dynamics is suppressed. For strong circle swimmer of large chirality the motion of particle is mostly confined to its own location and no clustering is observed. The interaction among the particle leads to such three distinct phases, whereas for noninteracting chiral system only two types of dynamics is observed. The presence of three  distinct phases for different chirality gives detail understanding of effect of chirality on the particles dynamics and steady state properties. This can be useful to use sorting of particles, based on their chirality. Hence our study can be useful in 
pharmaceutical industry as well as in clinical therapeutics and sorting of drugs\cite{drug,bechinger,nourhani2015guiding,mijalkov2013sorting}.

\section{conflict of interest}
Here is no conflict of interest.
\section{Acknowledgement}
 The authors gratefully acknowledge the DST support for funding this project. S. Mishra thanks DST, SERB (INDIA), Project No. ECR/2017/000659 for partial financial support.

%%%%%%%%%%%%%%%%%%%%%%%%%%%%%%%%%%%%%%%%%%%%%%%%%%%%%%%%%%%%%%%%%%%%%%%%%%%%%%%%%%%%
\bibliographystyle{apsrev4-1}
\bibliography{references}
 \end{document}